# Effect of composition on the dielectric relaxation of zeolite-conducting polyaniline blends

I. Sakellis, A.N. Papathanassiou (*) and J. Grammatikakis

University of Athens, Physics Department, Section of Solid State Physics, Panepistimiopolis, 15784 Zografos, Athens, Greece

**Abstract**

The complex permittivity of conducting polyaniline and zeolite - polyaniline blends was measured in the frequency range $10^{-2}$ to $2 \times 10^{6}$ Hz from room temperature to liquid nitrogen temperature. A loss peak is detected for 25, 35 and 50 wt % zeolite blends. Its position in the frequency domain, activation energy and intensity is a function of composition. The experimental results are interpreted through the competing multiple role of zeolite: as being less conducting than polyaniline, it impedes the electric charge flow by dividing large conducting polyaniline regions into smaller pieces, subsequently provides short-range pathways and, moreover, enriches the blend in interfaces separating zeolite from polyaniline.

(*) Corresponding author; E-mail address: antpapa@phys.uoa.gr

## 1. Introduction

Electric charge transport in conducting polymers has attracted theoretical and experimental investigations. The objectives are the construction of a generalized frame for electrical conductivity in conducting polymers [1-3] and the preparation of well characterized metallic or semi-conducting polymers for many different technological applications, such as solar cells, pH electrodes, media for hydrogen storage, electronic devices etc [4, 5]. Complex permittivity measurements are valuable for studying electric charge flow of different scale and can therefore probe the very local structure of disordered media [6-11]. Porous inorganic materials can host conducting polymers into their porosity yielding nano-composite materials [12]. Zeolite contains pores, channels and cages of different dimensions and shapes and their surface is negatively charge-balanced with exchangeable cations [5]. Polyaniline (PAni) may complex with zeolite in composites that are characterized by the fast electronic mobility of PAni and the capability of zeolite to incorporate cations into its structure. In the present paper the effect of composition of zeolite-PAni blends on the effective electric and dielectric behavior is investigated.

## 2. Experimental

Freshly distilled Polyaniline (PAni) (Merck AR) was polymerized in the presence of $FeCl_3$ as oxidant in hydrochloric acid - water solutions at pH = 2.00 in an ice bath. The molar ratio of oxidant to monomer was 1:1 and the solvent used was triply distilled water. Purified Linde Type-N (LTN) zeolite with cavity structure with pores of approximately 3.56 A in diameter, was used to prepare the zeolite-PAni polymer solutions. Precipitates were purified by Soxhlet extraction for 35 h [13]. Zeolite-PAni disc shaped specimens 13 mm in diameter and about 1.5 mm thick were made in a IR press.

The specimens were placed in a vacuum cryostat operating at 1 Pa. Good contact between the surfaces of the specimen and the electrodes was achieved by attaching silver paste. The temperature was monitored from room temperature to liquid nitrogen temperature. The measurements were performed in the frequency range from $10^{-2}$ to $2 \times 10^6$ Hz by a Solartron SI 1260 impedance analyzer.

## 3. Results and Discussion

The real and the imaginary part of the (relative) permittivity ($\varepsilon'$ and $\varepsilon''$) are connected to the real and the imaginary part of the complex conductivity ($\sigma'$ and $\sigma''$) respectively, by the following relations [14]:

$$\varepsilon'(\omega) = \varepsilon_\infty + \frac{\sigma''(\omega)}{\varepsilon_0 \omega} \tag{1}$$

$$\varepsilon''(\omega) = \frac{\sigma'(\omega)}{\varepsilon_0 \omega} = \frac{\sigma_0}{\varepsilon_0 \omega} + \varepsilon_d'' \tag{2}$$

where $\omega$ denotes the angular frequency ($\omega = 2\pi f$; f is the frequency), $\varepsilon_0$ is the permittivity of free space, $\varepsilon_\infty$ is the high frequency permittivity, $\sigma_0$ is the frequency-independent conductivity (usually labeled as the dc conductivity) and $\varepsilon_d''$ is the imaginary part of the permittivity after subtracting the dc component;

$$\varepsilon_d'' = \frac{\sigma'(\omega) - \sigma_0}{\varepsilon_0 \omega} \tag{3}$$

The complex permittivity was measured in the frequency domain (from $10^{-2}$ to $2 \times 10^6$ Hz) at various temperatures ranging from the LNT to room temperature. Isotherms of the measured imaginary part of the dielectric constant vs frequency at room temperature for PAni and zeolite-PAni blends are depicted in Figure 1. For PAni and 10 wt % zeolite-PAni sample, $\log \varepsilon''(\log f)$ is dominantly a straight line with slope equal to $-1$, indicating that the conductivity is frequency independent. The variation of the frequency-independent real part of the conductivity $\sigma_0$ as a function of composition is shown in the inset of Figure 1. A broad dielectric loss mechanism appears in the high-frequency region for zeolite-rich blends. Isotherms of the measured $\varepsilon''$ vs frequency for the 25, 35 and 50 wt % zeolite-PAni blends are shown in Figure 2. The dielectric loss $\varepsilon_d''(f)$ obtained after subtraction of the dc constituent are presented in Figure 2. The dominant broad relaxation peak shifts towards lower frequency, when temperature is reduced. The frequency $f_{max}$ where a relaxation peak has its maximum is plotted as a function of reciprocal temperature is presented in Figure 3. The Arrhenius law is employed to fit the experimental data:

$$f_{max} = f_0 \exp\left(-\frac{E}{kT}\right) \tag{4}$$

where k is the Boltzmann's constant, $f_0$ is a pre-exponential factor and E denotes the activation energy. The estimated activation energy values are cited in Table I. The values enlisted are more than one order of magnitude lower than those reported for dielectric relaxation in any type of zeolite, but are comparable with activation energy values of relaxation in conducting polymers [10]. It is therefore likely that relaxation involves the short-range polaron motion. The intensity of the relaxation $\Delta\varepsilon$ as a function of reciprocal temperature follows the $T^{-1}$ Curie law (Figure 4).

Barton, Nakajima and Namikawa [15-17] suggested that the position of the relaxation mechanism be determined by the dc conductivity, through the so-called BNN condition:

$$f_{max,BNN} = \frac{1}{2\pi p} \frac{\sigma_0}{\varepsilon_0 \Delta\varepsilon} \tag{5}$$

where p is a constant of the order of unity. In Figure 5, the experimental values of $f_{max,BNN}$ are displayed vs the experimentally observed $f_{max}$. Significant discrepancy occurs between experimental results and theory. Disagreement has been observed in some amorphous materials. They were attributed to over-simplifications of the theoretical model, the material's characteristics and the nature of charge carriers are ignored, a fixed hopping mechanism is assumed, dipole relaxation and many body long range interaction are ignored etc. [7, 9].

The increase of zeolite content affects the relaxation peak in three ways:
(i) $f_{max}$ tends to shift towards low frequency.
(ii) E has a distinct maximum for the 35 wt % zeolite blend.
(iii) $\Delta\varepsilon$ is significantly suppressed for the 35 wt % zeolite blend.
Zeolite, as being less conducting than PAni does, prohibits the electric charge flow, induces additional disorder by setting spatially randomly distributed energy barriers and provides interfaces separating zeolite from PAni. For conventional reasons, zeolite and PAni will be mentioned as the insulating and conducting phase, respectively. The dielectric properties in disordered media result from spatial and energy disorder; they both determine boundary conditions to the electric charge flow and yield capacitive phenomena. In conducting polymers, for example, dead-end polymer chains, orientation disorder of polymer chains and inhomogeneous disorder stemming from the existence of polaron-rich territories (conducting grains) into a less conducting environment [18-21], give rise to capacitive effects, which –in the frequency domain- appear as dielectric relaxation. In the ac response the charge carriers oscillate between adjacent potential wells, though in the dc conductivity have to move a distance comparable to the sample dimensions to produce a macroscopically measurable result. With increasing doping with zeolite, the polymer conducting grains, among other distresses, suffer division in smaller pieces, which prevents carriers from long-range displacement, however giving them the possibility to contribute to ac conductivity. Within this frame, $\sigma_0$ decreases on increasing the zeolite fraction. On the other hand, the environment around relaxing charges along short-range lengths is perturbed by the augmentation of the insulating phase; as mentioned above, the increase of zeolite content induces additional boundary conditions to the short-range motion of polarons and, therefore, the relaxation process requires longer relaxation times on increasing the insulating phase and, therefore, $f_{max}$ shifts to lower frequency on increasing the zeolite content.

Interfacial polarization phenomena in binary systems consisting of materials with different electrical conductivity and (static) permittivity values were treated a long time ago [22]. An overview for dielectric properties of nanodispersion or of porous media can be found in Refs [23, 24]. In general, models predicting the position of the interfacial dielectric relaxation and its dependence on composition, are based on two arbitrary assumptions: (i) the host material is homogeneous and (ii) the inclusions have a certain geometry and shape. As a result, qualitative predictions are successful in simple model systems. For the blends studied in the present paper, it is quite abrupt to use a specific model and compare the experimental results with theoretical predictions. However, we can have some rough estimate. Adopting the simple case of two-layer system, the relaxation time $\tau$ is:

$$\tau = \frac{\varepsilon_1 d_2 + \varepsilon_2 d_1}{\sigma_1 d_2 + \sigma_2 d_1} = \frac{\varepsilon_1 + \varepsilon_2 (d_1/d_2)}{\sigma_1 + \sigma_2 (d_1/d_2)} \tag{6}$$

where the subscripts 1 and 2 indicate phase 1 (PAni) and 2 (zeolite), respectively and d is the thickness of an individual phase. The thickness ratio can be replaced by the volume ratio, for the specific two-layer system, since the two phases share the same interfacial surface area. However, since the latter model is used to interpret *qualitatively* our results, we have not made volume ratio calculations. From the low frequency data of our measurements at room temperature, we have $\varepsilon_1 \approx 10^4$, $\sigma_1 \approx 0.2 \, S/m$. For zeolite we take a couple of typical values from the literature $\varepsilon_2 \approx 40$, $\sigma_2 \approx 10^{-7} \, S/m$ [25]. By substituting these values into Eq. (6), we get $\tau \approx 10^{-5} \, s$, which yields a maximum of a loss beak around $f_{max} = \tau^{-1} \approx 10^5 \, Hz$. Moreover, the increase of zeolite fraction yields a very weak shift of $f_{max}$ toward lower frequency (e.g., less than 0.5%, for the available composition blends). We worked on the two-layer model for reasons of simplicity; note that similar behavior is predicted if the blend is approximated as a conducting matrix accommodating spherical insulating inclusions (Wagner's approximation). As can be seen in Figure 3, the predicted $f_{max}$ value (at room temperature) is roughly of the same order of magnitude with the experimental values. However, although $f_{max}$ is reduced on increasing the zeolite fraction, in accordance with the qualitative prediction of Eq. (6), the shift detected experimentally is roughly about one order of magnitude, which is significantly different from the model prediction. These findings support the idea that the broad relaxation might be related with polarization processes related with the interfacial interaction of zeolite and PAni, but the process is much more complicated than that described by emulsion science for simple mixtures. In our case, electric charge flow in PAni itself is a quasi-1D process within an inhomogeneously disordered polymer environment. On the other hand, zeolite is porous material. Not only are the different phases structurally complicated, but also the interface will subsequently not well defined.

One would expect that the increase of the insulating phase (zeolite) merely result in a systematic increase of the activation energy. However, E takes a maximum value for 35 wt % zeolite. At the same composition, the loss strength is suppressed. At low concentration of zeolite, its dominant role seems to be the separation of conducting PAni regions in smaller pieces. In this way, electric charge flow is impeded, reflecting in the increase of the activation energy from 25 to 35 wt % zeolite. Zeolite interrupts long conducting pathways: for this reason the dc component gets weaken and, at the same time, the number of small-dimension conducting islands increases (as a result of division of larger conducting territories from the presence of insulating zeolite regions), feeding the ac response with additional pathways for short range polaron transport. By further increasing the zeolite content (i.e., reaching the 50 wt % composition) the blend becomes rich enough in interfaces offering additional low resistance grain boundary pathways. The result of the competing mechanisms described above, is the appearance of a maximum in E and a minimum Δε at an intermediate composition (35 wt % zeolite).

**4. Conclusions**

Frequency domain complex permittivity studies in conducting PAni and 10, 25, 35 and 50 wt % zeolite blends showed that a dielectric loss peak appears for 25, 30 and 50 wt % zeolite (in the frequency window $10^{-2}$ - $2 \times 10^6$ Hz). The activation energy values for relaxation are typical of polaron relaxation. The location of the peak, the activation energy and intensity (at any given temperature) is a function of the zeolite content. A maximum in the activation energy values and a corresponding minimum to the loss intensity is interpreted through the contradicting multiple role of zeolite: as being less conducting than PAni, it impeds the electric charge flow by dividing large conducting PAni regions into smaller pieces, subsequently provides short-range pathways and, moreover, enriches the blend with interfaces separating zeolite from PAni.

**Acknowledgements**

We thank S. Sakkopoulos, E Vitoratos and E. Dalas (University of Patras) for providing us with the specimens studied in the present work.

**Table I.** Activation energy values for dielectric relaxation.

**Table I**

| % wt zeolite | E (eV) |
|---|---|
| 25 | 0.0246 ± 0.0007 |
| 35 | 0.075 ± 0.005 |
| 50 | 0.0317 ± 0.0007 |

**Figure Captions**

**Figure 1:** Isotherms of the imaginary part of the permittivity vs frequency at room temperature (292K) for conducting PAni and zeolite-PAni blends (curves from top to bottom correspond to PAni, 10, 25, 35 and 50 wt % zeolite-PAni blends). Inset: Room temperature frequency independent conductivity $\sigma_0$ vs the wt % content in zeolite.

**Figure 2:** Temperature dependence of the measured $\varepsilon''$ (first column) and $\varepsilon_d''$ obtained after subtraction of the dc component from the measured $\varepsilon''$ (last column) vs frequency. In each plot, curves from top to bottom correspond to 292 K, 248 K, 175 K, 135 K, 107 K, 87 K.

**Figure 3:** $f_{max}$ against reciprocal temperature. Squares: 25 wt % zeolite Circles: 35 wt % zeolite; Triangles: 50 wt % zeolite. Straight lines are best fits of the Arrhenius law to the data points.

**Figure 4:** The intensity of the dielectric loss mechanism $\Delta\varepsilon$ against reciprocal temperature. Squares: 25 wt % zeolite Circles: 35 wt % zeolite; Triangles: 50 wt % zeolite.

**Figure 5:** The maximum loss frequency $f_{max,BNN}$ obtained from Eq. (5) vs the measured $f_{max}$. Squares: 25 wt % zeolite Circles: 35 wt % zeolite; Triangles: 50 wt % zeolite. The straight line is predicted by the empirical BNN model, assuming that the parameter p=1 (Eq. (5)). Squares: 25 wt % zeolite Circles: 35 wt % zeolite; Triangles: 50 wt % zeolite.

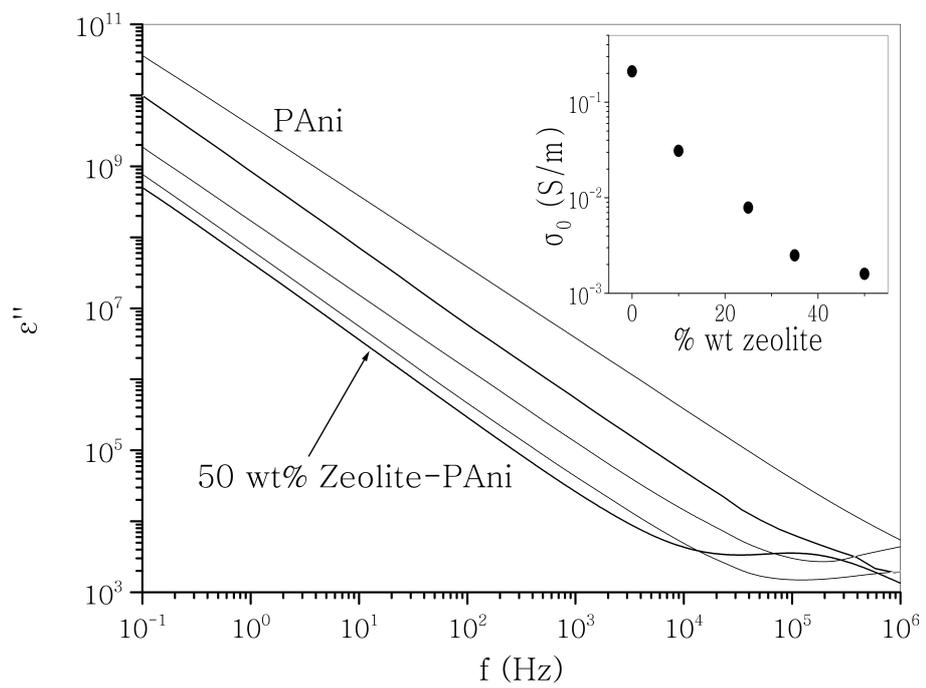

FIGURE 1

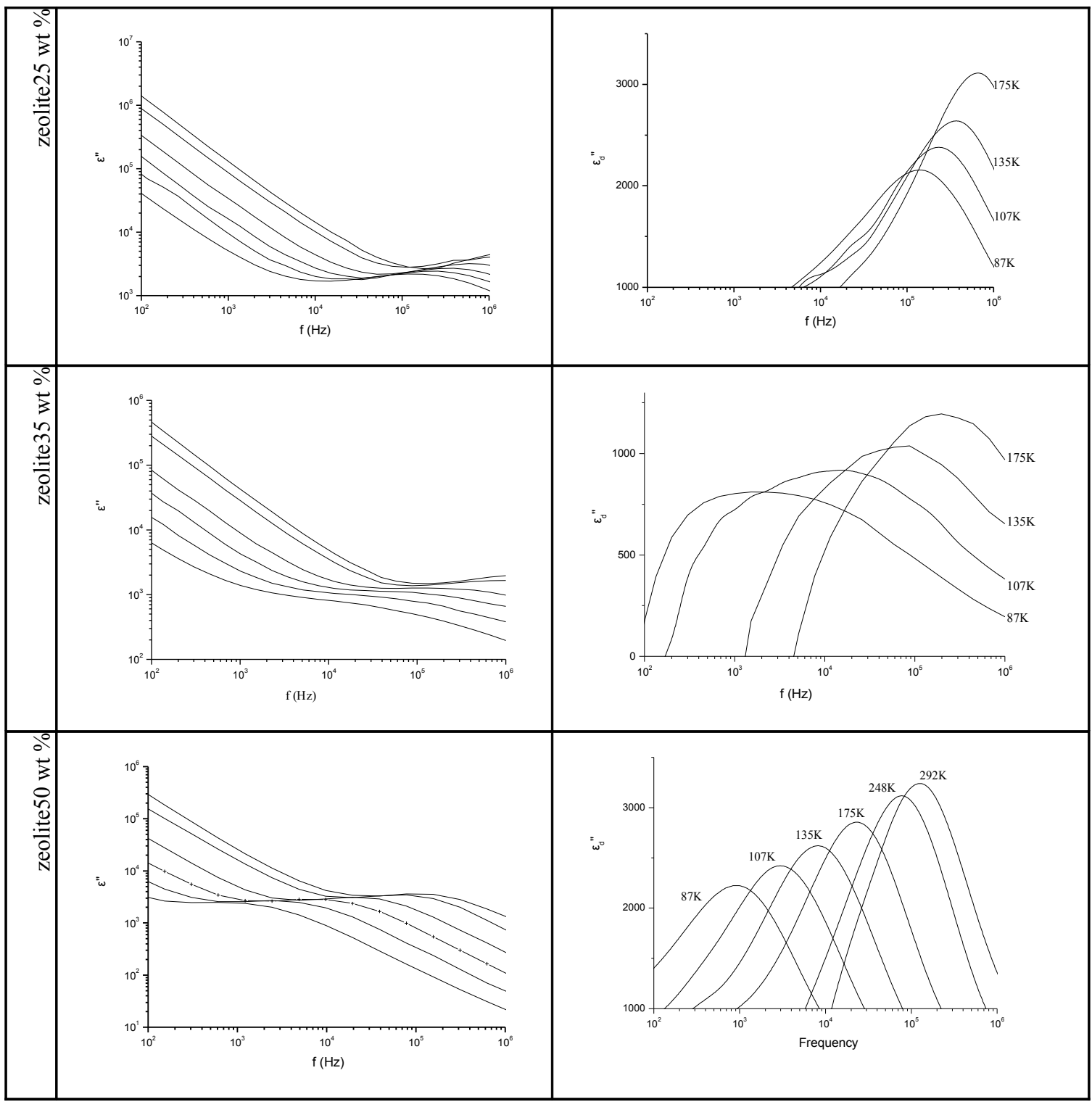

FIGURE 2

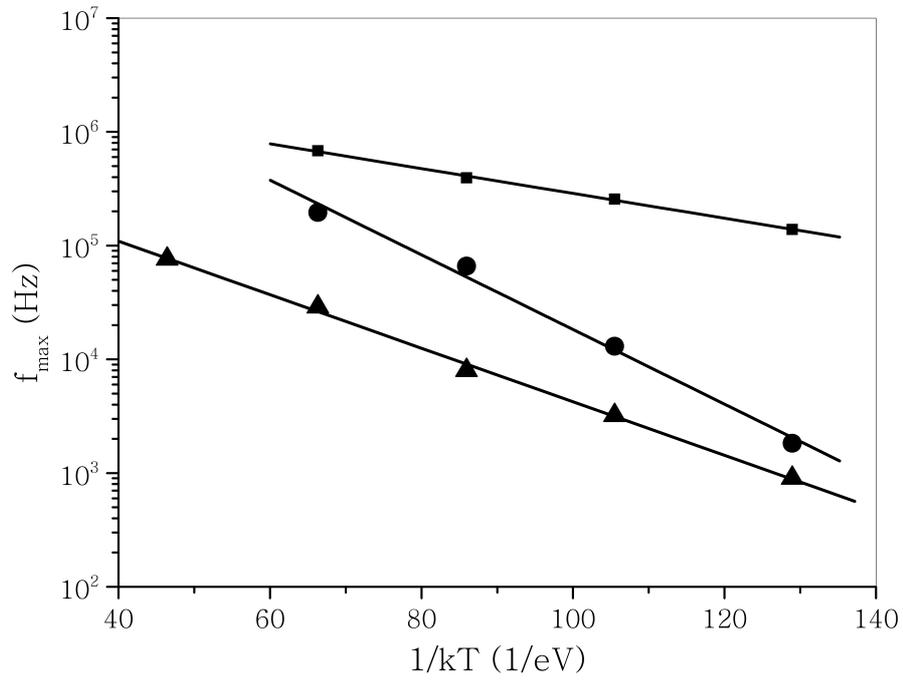

FIGURE 3

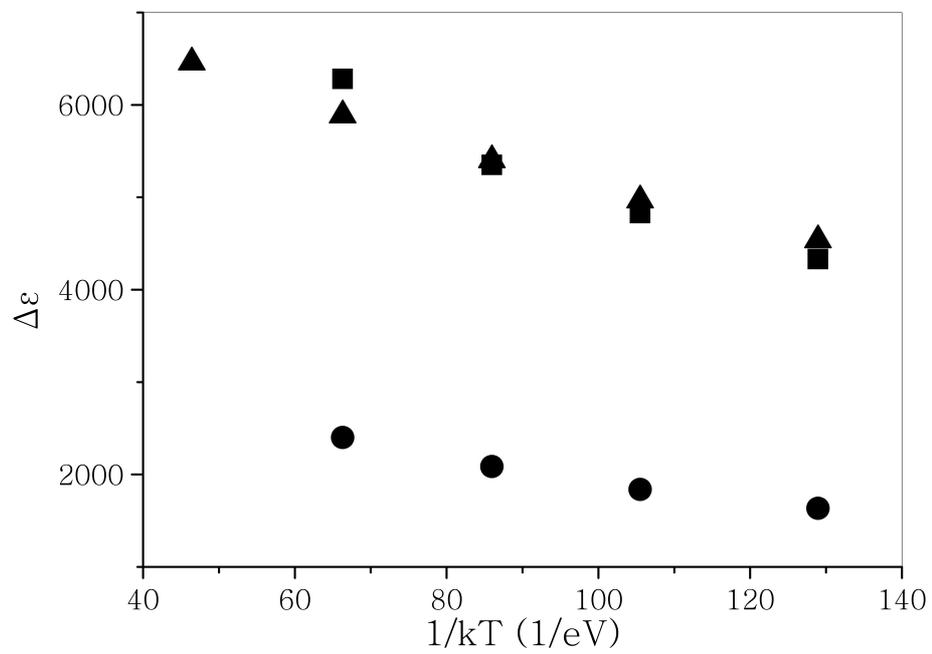

FIGURE 4

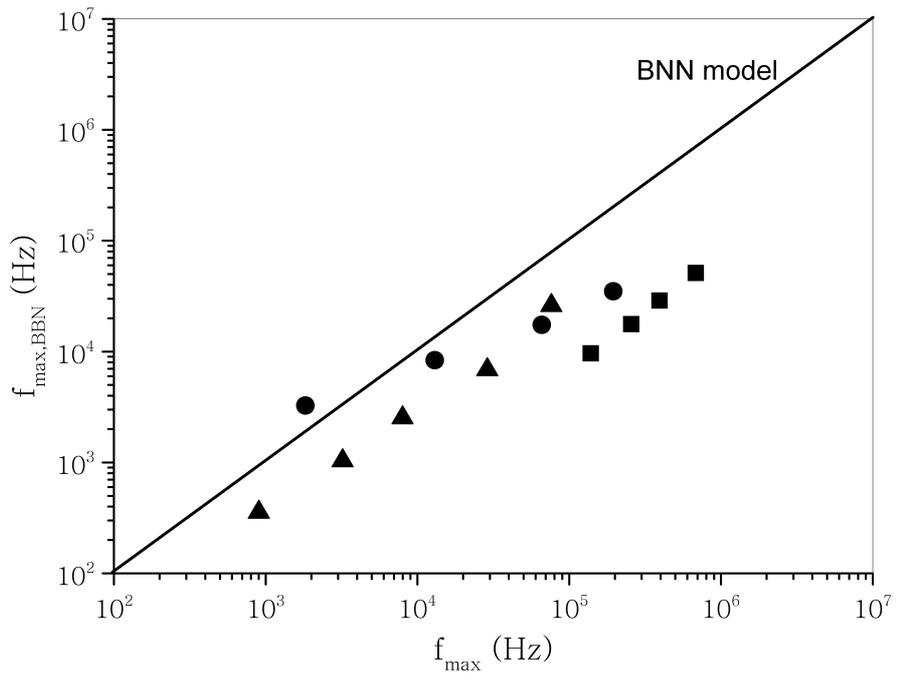

FIGURE 5